\documentclass[10pt,journal]{IEEEtran}
\usepackage{amsmath,amsfonts}
\usepackage{algpseudocode}
\usepackage{algorithm}
\usepackage{array}
\usepackage[caption=false,font=normalsize,labelfont=sf,textfont=sf]{subfig}
\usepackage{textcomp}
\usepackage{stfloats}
\usepackage{url} 
\usepackage{verbatim}
\usepackage{graphicx}
\usepackage{cite}
\usepackage{multirow}
\usepackage{wasysym}
\usepackage{authblk}
\usepackage{enumitem}
\usepackage{tcolorbox}

\usepackage{xcolor, colortbl}

\usepackage{accents}
\newlength{\dhatheight}
\newcommand{\doublehat}[1]{%
    \settoheight{\dhatheight}{\ensuremath{\hat{#1}}}%
    \addtolength{\dhatheight}{-0.35ex}%
    \hat{\vphantom{\rule{1pt}{\dhatheight}}%
    \smash{\hat{#1}}}}
\newcommand*{\dbar}[1]{\overline{\overline{#1}}}

\begin{document}
\bstctlcite{IEEEexample:BSTcontrol}

\title{Note-Level Singing Melody Transcription for Time-Aligned Musical Score Generation}

\author{Leekyung Kim, Sungwook Jeon, Wan Heo, Jonghun Park \IEEEmembership{}
\thanks{Manuscript received 27 August 2021; 
\textit{(Corresponding author: Jonghun Park.)}}
\thanks{The authors are with the Department of Industrial Engineering and Institute for Industrial Systems Innovation, Seoul National University, Seoul, Republic of Korea(e-mail: klk97@snu.ac.kr; wookee3@snu.ac.kr; jrw8217@snu.ac.kr; jonghun@snu.ac.kr).}
}

\markboth{IEEE/ACM Transactions on Audio, Speech, and Language Processing
}{LeeKyung Kim, Sungwook Jeon, \MakeLowercase{\textit{(et al.)}}: Note-Level Singing Melody Transcription for Time-Aligned Musical Score Generation}

\IEEEpubid{0000--0000/00\$00.00~\copyright~2024 IEEE}

\maketitle

\begin{abstract}
Automatic music transcription converts audio recordings into symbolic representations, facilitating music analysis, retrieval, and generation. A musical note is characterized by pitch, onset, and offset in an audio domain, whereas it is defined in terms of pitch and note value in a musical score domain. A time-aligned score, derived from timing information along with pitch and note value, allows matching a part of the score with the corresponding part of the music audio, enabling various applications.
In this paper, we consider an extended version of the traditional note-level transcription task that recognizes onset, offset, and pitch, through including extraction of additional note value to generate a time-aligned score from an audio input. To address this new challenge, we propose an end-to-end framework that integrates recognition of the note value, pitch, and temporal information. This approach avoids error accumulation inherent in multi-stage methods and enhances accuracy through mutual reinforcement. Our framework employs tokenized representations specifically targeted for this task, through incorporating note value information.
Furthermore, we introduce a pseudo-labeling technique to address a scarcity problem of annotated note value data. This technique produces approximate note value labels from existing datasets for the traditional note-level transcription. Experimental results demonstrate the superior performance of the proposed model in note-level transcription tasks when compared to existing state-of-the-art approaches. We also introduce new evaluation metrics that assess both temporal and note value aspects to demonstrate the robustness of the model. Moreover, qualitative assessments via visualized musical scores confirmed the effectiveness of our model in capturing the note values.
\end{abstract}

\begin{IEEEkeywords}
automatic music transcription, music information retrieval, singing melody transcription, note value detection
\end{IEEEkeywords}

\section{Introduction}

\begin{figure}[!t]
    \centering
    \includegraphics[width=\linewidth]{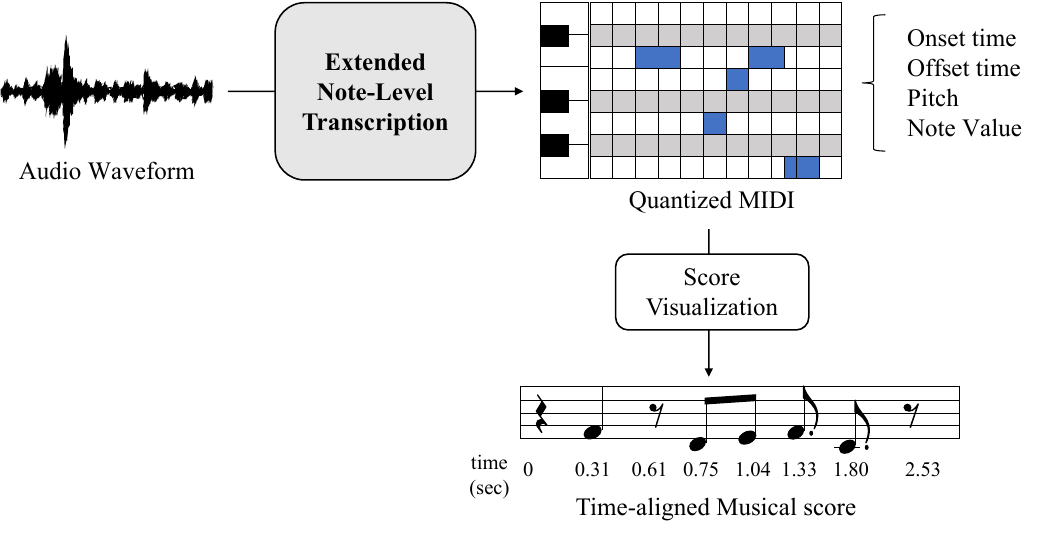}
    \caption{An overview of the extended note-level singing melody transcription task proposed in this paper.}
    \label{fig_ov}
\end{figure}

\IEEEPARstart{A}{utomatic} Music Transcription (AMT) is a task of converting musical audio into a form of music notation. It plays a crucial role in various applications such as music analysis, search, and generation. Singing melody transcription, a subtask of AMT, specifically aims to recognize music notation for vocal melodies from polyphonic audio recordings. It holds significant importance in the field of Music Information Retrieval (MIR) due to the distinctive role of melody, mainly performed by singers in pop music, as a recognizable aspect that distinguishes one composition from another.

Each task of AMT recognizes various musical features and notations, including fundamental frequency (F0)\cite{f0} curve at each frame and piano-roll\cite{klapuri}. Among these, a musical score, which visually represents a recording in terms of musical symbols, is one of the final outputs required by AMT. Note is a fundamental symbol in music notations that facilitates clear communication and interpretation of musical ideas. It is represented with quantized pitch and note value in a musical score. The pitch and note value indicate the perceived frequency and the relative length of a note, respectively. For example, if we consider a quarter note as the unit of note value, an eighth note (\eighthnote) is said to have a note value of 0.5 since its length is half of that of a quarter note (\quarternote). 
 
\IEEEpubidadjcol
On the other hand, note-level transcription task characterizes notes in terms of the onset and offset times that respectively refer to the beginning and end of a note, along with the pitch. The output of the note-level transcription task corresponds to a piano-roll representation that serves as an intermediate goal towards recognizing a musical score. It should be noted that onset and offset times are represented by continuous time units in the musical audio domain, while note value in a musical score is represented in terms of the discrete beat units.

A time-aligned score enriches the musical score through incorporating onset and offset times of each note. This alignment enables multi-modal navigation of music by synchronizing each note in the score with its occurrence time in the music audio. Such synchronization facilitates various applications. These include verification and manual error correction of transcription results, data collection for audio-to-score alignment, tempo and rhythm analysis, and synchronized visualization of the transcribed score, which is beneficial for karaoke systems and musical instrument tutoring.

Despite its applications, there is scarcity of approaches that directly extract the time-aligned scores from a piece of musical audio. Relevant approaches are listed in Table \ref{related}, along with their differences from our approach, which will be referred to as an extended note-level transcription. Audio-to-symbolic transcription\cite{weakly_attention, beat_attention, CRNN_HSMM, ctc_score1, ctc_score2} extracts a musical score that includes pitch and note value, without time information.
Audio-to-score alignment\cite{a2s-eu, a2s-ie} produces a time-aligned score, but it requires a musical score as an input with a musical recording. Existing methods to obtain a time-aligned score from a musical audio file often rely on quantization\cite{quantization} after the note-level transcription, where the quantization process converts the onset and offset times into the note values\cite{midi_to_score}. 
However, these methods were not designed for direct recognition of time-aligned scores and suffer from inherent limitations.

\begin{table*}
\centering
\caption{Comparisons of relevant tasks for time-aligned scores}
\label{related}
\begin{tabular}{l|l|l|l}
\hline
Task                                       & Input                     & Output                                                                                    & Difference       \\ \hline
Note-level transcription\cite{klapuri}                   & audio                     & piano-roll                                                         & output without note value           \\
Quantization\cite{quantization}                               & piano-roll  & musical score (implicitly time-aligned by input) & input            \\
Audio-to-symbolic transcription\cite{weakly_attention, beat_attention, CRNN_HSMM, ctc_score1, ctc_score2}            & audio      & musical score       & output without time\\
Audio-to-score alignment\cite{a2s-eu, a2s-ie}                   & audio, musical score      & time-aligned musical score                                                                        & additional input \\
\textbf{Extended note-level transcription (proposed)} & \textbf{audio}            & \textbf{time-aligned musical score}                                                               & \textbf{-}       \\ \hline
\end{tabular}
\end{table*}

Motivated by these challenges, we attempt to extend the conventional note-level transcription with the goal of directly extracting a time-aligned score from musical audio. Specifically, in addition to recognizing the onset time, offset time, and pitch of a note as in the note-level transcription, the note value is recognized. With this information, a time-aligned score can be obtained under the assumption of a fixed time signature. We will refer to this task as the extended note-level transcription throughout this paper. Fig. \ref{fig_ov} provides an overview of the extended note-level singing melody transcription task.

Since the prior work addressed the estimation of timing and note values as independent tasks, they face significant challenges, including the need for separate training and inference at each stage and the error accumulation across stages. In contrast, the proposed unified end-to-end pipeline simplifies the transcription process, eliminates the multi-stage dependencies, and enhances practical applicability by directly generating time-aligned scores.

In the proposed framework, the singing melody is represented as a monophonic sequence of pre-defined tokens, predicted by a deep neural network based on an input spectrogram. In this paper, a new token set is defined to encode onset, offset, pitch, and note value information.

To address the limited availability of annotated note value labels, we propose a pseudo-labeling method that generates note values from existing note-level transcription datasets, such as ST500\cite{st500} and DALI\cite{dali}, which lack time-aligned note value labels. While HSD\cite{hsd} appears to be the only available dataset with the time-aligned note value labels for the considered task, its limited size of 68 songs is insufficient to train a model using supervised learning in most cases. Through training the model with these pseudo-labeled data, we demonstrate the usefulness of the proposed pseudo-labeling method for model training.

Furthermore, existing evaluation metrics commonly used in the existing note-level transcription research\cite{st500, jong, musicyolo} do not account for the note values. 
Although audio-to-score alignment tasks\cite{a2s-eu, a2s-ie} produce a time-aligned musical score, their evaluation metrics focus solely on assessing the predicted onset and offset times for a given note sequence, making them unsuitable for evaluating time-aligned note values. 
To address this limitation, we propose new evaluation metrics tailored to assess the recognition of time-aligned note values, an additional feature considered in this study. These metrics allow comprehensive evaluations of performances using the HSD dataset.

Our contributions are summarized as follows:
\begin{itemize}[noitemsep]
    \item We propose an end-to-end framework for the new task named an extended note-level singing melody transcription, enabling direct generation of time-aligned scores by integrating onset, offset, pitch, and note value recognition into a unified process.
    \item To address the scarcity of annotated time-aligned note value data, we propose a pseudo-labeling method that generates note value labels by using the existing note-level transcription datasets with beat tracking results.
    \item We define a new task-specific token set that encodes note onset, offset, pitch, as well as note value information and utilize Transformer model to predict the token sequences.
    \item Novel evaluation metrics for time-aligned note value recognition, adapted from conventional note-level transcription evaluation metrics, are introduced to provide comprehensive assessment of recognition performance. 
    \item Extensive experiments show that the proposed model improved the note-level, symbolic, and time-aligned note value transcription performance compared to prior methods.
\end{itemize}

\section{Related Work}

\subsection{Note-Level Singing Melody Transcription}
Note-level singing melody transcription refers to a task of recognizing onset time, offset time, and pitch of a monophonic vocal melody from a polyphonic music recording. Traditional methods typically rely on a multi-stage approach, including note segmentation after F0 estimation\cite{jdc} and transcription after pre-processing the input\cite{st500}. This multi-stage approach is prone to error propagation and requires individual training and inference for each stage.

Currently, the state-of-the-art performance for the note-level singing melody transcription can be found in note-level Transformer\cite{jong} and MusicYOLO\cite{musicyolo}. 
These models employ an end-to-end framework that has a single model to obtain note-level annotations from music audio signals, thereby overcoming the limitations of multi-stage methods. Yet, MusicYOLO requires an additional vocal separation model\cite{spleeter} for singing melody transcription from multitrack music, as it focuses on transcribing from monophonic human voice.

Recent singing voice transcription (SVS) research\cite{av-svt, altamt, rosvot} has addressed the singing melody transcription task through incorporating additional input modalities. \cite{av-svt, altamt} leveraged self-supervised pre-training models\cite{wav2vec, av-hubert} and multi-modal learning techniques that utilized both audio and video inputs. ROSVOT\cite{rosvot} incorporated word boundaries to align note sequences with corresponding word sequences with the goal of establishing an annotation pipeline for training an SVS model.

Although deep learning models have achieved good performances in the note-level singing melody transcription, post-processing is still necessary to quantize the transcription result into a musical score. Beat quantization typically involves grid quantization\cite{quantization}, which converts the time to the nearest beat grid, or rhythm quantization, where a trained model\cite{nakamura_rhythm} converts time into note values. Such an additional step results in a multi-stage process, whereas the proposed framework achieves this directly in a fully end-to-end manner.

\subsection{Audio-to-Symbolic Transcription}
Audio-to-symbolic transcription involves recognizing the pitch and note value of musical notes within an audio recording for generating a musical score.
Previous research \cite{weakly_attention, beat_attention, CRNN_HSMM, ctc_score1, ctc_score2} has primarily focused on transcribing vocal melodies to generate lead sheets that are streamlined musical notations focusing on the melody. This task shares similarities with the objective of this paper in terms of extracting the pitch and note value of a vocal melody.

To predict pitch and note value directly from an audio input, previous research has employed encoder-decoder models with an attention mechanism\cite{weakly_attention, beat_attention}, encoder-decoder models with CTC loss\cite{ctc_score1, ctc_score2}, or hybrid models that combine a musical language model with an acoustic model\cite{CRNN_HSMM}. Yet, these methods require additional steps such as vocal separation\cite{ctc_score1, ctc_score2}, beat tracking\cite{weakly_attention, beat_attention, CRNN_HSMM}, or a unified tempo across input audio\cite{weakly_attention} as they were trained on constrained input. These constraints limit their applicability in real-world scenarios.

The key distinction of this paper lies in the additional recognition of the onset and offset times for the time-aligned scores. Approximating these times from note values requires additional information such as tempo, beat information in the audio, and the length of regions without vocal melodies. Roughly estimated time with this additional information is not accurate enough for tasks like multi-modal navigation. 

While symbolic musical scores provide a structured representation of pitch, rhythm, and note values, they pose limitations on capturing the timing variations introduced by a performer's expressive style. For instance, singers may intentionally extend or shorten notes relative to their written values as a part of their unique interpretation. In contrast, note-level transcription captures the actual timing of a performance but lacks the structured representation. Accordingly, a time-aligned musical score that integrates both symbolic and note-level transcription results is essential to bridge this gap.
This integrated representation allows for comprehensive understanding of both musical structure and performance style, enabling applications such as performance analysis, cover song comparison, and music education. These practical applications highlight the broader impact and significance of our work. 

\subsection{Sequence-to-Sequence Method}
Sequence-to-sequence method\cite{seq2seq} has been utilized to predict a target sequence from an input sequence, with an adaptability to varying sequence lengths. This approach has demonstrated robust performance across various domains\cite{s2s_amt, s2s_asr} including transcription in the MIR field. 

Automatic Singing Transcription (AST) was initially defined as a sequence-to-sequence problem in \cite{cs2s}, where an input sequence, consisting of audio spectrograms at the frame level, is translated into a sequence of musical symbols. 
Transformer\cite{transformer} with attention mechanisms has emerged as a popular model for the sequence-to-sequence learning that employs an autoregressive prediction of an output sequence.

\subsection{Semi-Supervised Learning}
Semi-supervised learning (SSL) is a widely used method to address the issue of insufficient labeled data through leveraging large amounts of unlabeled data in conjunction with labeled data. 
In MIR, data scarcity presents a persistent challenge, and obtaining labeled data often requires considerable effort and cost to ensure accurate labeling. Many studies utilized SSL approaches including VAT\cite{vat}, CTC loss\cite{ctc}, and a pre-trained model\cite{jukebox} for transcription tasks\cite{vocano, ctc_note, notation1, notation2, notation3}.

Pseudo-labeling is another SSL approach that generates pseudo-labels for unlabeled data to facilitate learning. In a teacher\textendash student framework, a teacher model trained on a small amount of labeled data generates pseudo-labels for a larger amount of unlabeled data, and they are used by a student model for learning. This framework has been applied in various transcription studies\cite{adt, ssl_vocal,jdc_note}.

\section{Method}
The goal of the extended note-level singing melody transcription is to extract a token sequence representing a monophonic vocal melody in terms of onset, offset, pitch, and note value from polyphonic music. Spectrogram is adopted to represent audio features. Given input spectrogram $X_{spec}\in \mathbb{R}^{N_T\times N_F}$, where $N_T$ is the number of time frames and $N_F$ is the number of frequency bins, transcription system outputs the token sequence from fixed token set $\Sigma = \mathbb{T} \cup \mathbb{P} \cup \mathbb{V} \cup \mathbb{S}$ that represents note information of a singing melody. $\mathbb{T}$, $\mathbb{P}$, $\mathbb{V}$, and $\mathbb{S}$ are the sets of time tokens, pitch tokens, note value tokens, and special tokens, respectively.

Specifically, note $y_i$ is defined as a quadruple comprising onset time $\hat{t}_{i}\in \mathbb{T}$, offset time $\doublehat{t}_{i} \in \mathbb{T}$, pitch $p_i \in \mathbb{P}$, and note value $v_i \in \mathbb{V}$, where $i$ denotes the note index within a note sequence. $\mathbb{T}$ is a set given by $\mathbb{T}=\{0, \Delta, 2\Delta, \dots, T\}$, where $T$ is the length of the input audio in seconds, and $\Delta$ refers to the length of a time frame in seconds. $\mathbb{P}=\{0,1,\dots,127\}$ corresponds to MIDI pitch numbers, and $\mathbb{V}=\{1,2,\dots, v_{max}\}$, where $v_{max}$ indicates the maximum note value. The onset and offset times must satisfy the monophonic condition: $\hat{t}_{i} < \doublehat{t}_{i} \leq \hat{t}_{j} < \doublehat{t}_{j}$ if $i<j$, where $i$ and $j$ represent the note indices in a note sequence. Sequence of $N$ notes, $y$, is represented as a sequence of quadruples of tokens $(\sigma \in \Sigma)$ given by 
\begin{gather} \label{eq1}
y = \left[y_1, \dots, y_N\right] = \left[(\hat{t}_{1}, p_1, \doublehat{t}_{1}, v_1), \dots, (\hat{t}_{N}, p_N, \doublehat{t}_{N}, v_N) \right]
\end{gather}

The proposed transcription method aims to find the most likely note sequence, calculated as the product of the conditional probability that a token will appear, given the spectrogram and the previous tokens. That is:
\begin{gather}
\hat{y}=\arg\max_{y\in\Sigma^*}P\left(y|X_{spec}\right),
\label{eq2} \\
P\left(y|X_{spec}\right)=\prod_{i=1}^{4N}P\left(\sigma_i|\sigma_0,\dots,\sigma_{i-1},X_{spec}\right),
\label{eq3}
\end{gather}
where $\sigma_i$ is the $i^{th}$ token in a token sequence.

\begin{figure*}[!t]
    \centering
    \includegraphics[width=5.7in]{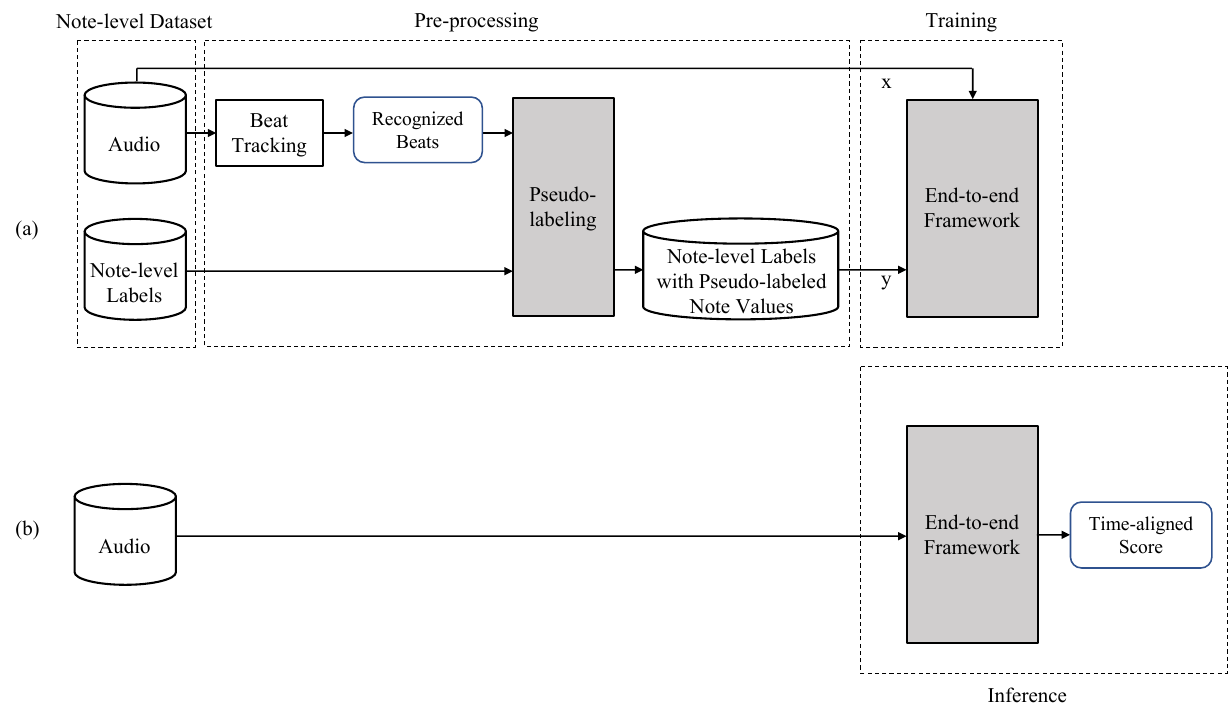}
    \caption{Proposed framework. (a) Note values were pseudo-labeled using the labels of the dataset for the note-level transcription and the beat tracking results obtained from an open-source library. Subsequently, an end-to-end model was trained using the audio along with the labels from the dataset for the note-level transcription incorporating the pseudo-labeled note values. (b) The trained model is capable of recognizing the onset, offset, pitch, and note value at once during inference. These recognition results enable the generation of a time-aligned score from an audio input directly.}
    \label{fig_framework}
\end{figure*}

Proposed framework illustrated in Fig. \ref{fig_framework} recognizes the onset and offset times, pitch, and note value from an audio signal, using the datasets for the extended note-level transcription.
Specifically, a Transformer encoder-decoder model is trained using a sequence-to-sequence method. In the inference phase, the trained model makes an autoregressive prediction of a token sequence based on an audio spectrogram.

\subsection{Pseudo-Labels From Dataset for Note-Level Transcription}
\label{section_pseudo}
Training a model to extract the onset, offset, pitch, and note value at the same time requires comprehensive labels for all four types of information along with their corresponding audio file. 
Unfortunately, the datasets containing such detailed information are limited in quantity. 
Accordingly, we propose a method for generating pseudo-labels from the datasets for the note-level transcription with a relatively large quantity.

The proposed pseudo-labeling method approximates the note value from the onset and offset labels in the note-level transcription datasets using grid quantization\cite{quantization}. 
Specifically, we use the madmom\cite{madmom_model} to detect beats in the input audio. It assumes a consistent time signature of either 3/4 or 4/4 throughout the song. The beat grids in the sixteenth note units are then derived by equally dividing the recognized beat intervals, as beats are predicted in the quarter note units. Subsequently, the onset and offset times are then quantized to the nearest sixteenth beat grid to determine the note values in the sixteenth note units. 

It should be noted that the beat detection is employed during the pseudo-labeling process, but not in the inference phase. It serves as a pre-processing step to generate approximate note value labels, enabling the model to learn without additional labeled datasets. During the inference, the proposed framework directly predicts onset time, offset time, pitch, and note value from the audio input without carrying out the pre-processing steps, enabling a full end-to-end system.

\begin{figure}[!t]
    \centering
    \includegraphics[width=0.8\linewidth]{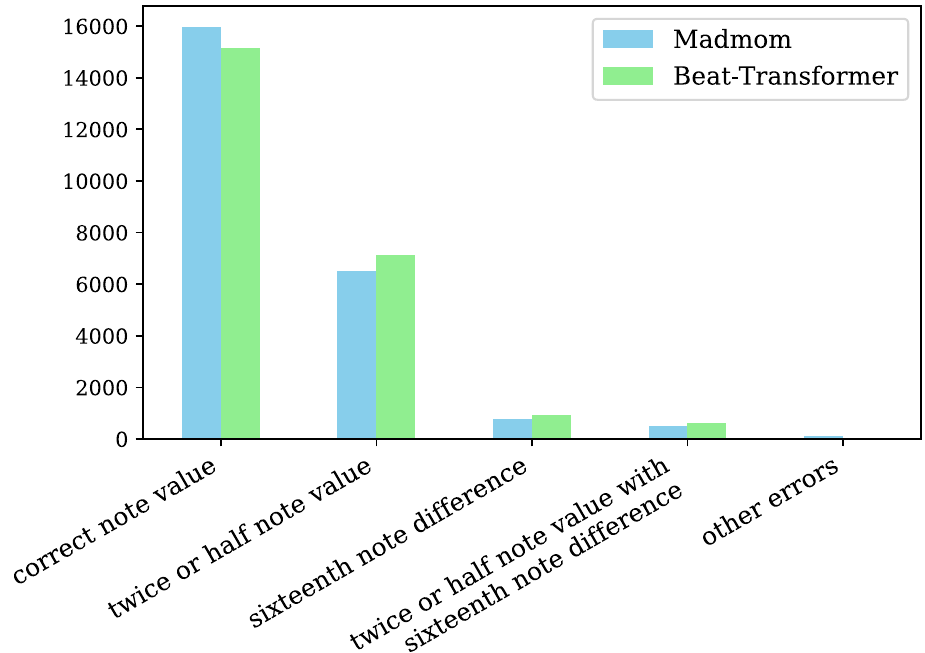}
    \caption{Distribution of pseudo-labeling outcomes using madmom and Beat-Transformer.}
    \label{fig_quan_err}
\end{figure}

To select the beat tracking model, we compared madmom with Beat-Transformer\cite{beat-transformer}, which offers the best performance among publicly available models under the same time signature assumptions. Fig. \ref{fig_quan_err} illustrates the distribution of pseudo-labeling outcomes for both models. This comparison involved applying the proposed pseudo-labeling method to the onset and offset times for the HSD dataset and evaluating the results against their ground-truth note value labels. Although both models exhibited reliable performance, madmom was selected due to its superior pseudo-labeling results.

The proposed note value labeling method still has limitations due to inaccurate beat detection results by madmom and inability to represent triplets, which cannot be expressed in terms of the sixteenth note units. Nevertheless, experimental results indicate that the note value recognition using our proposed pseudo-labeling method outperformed the other alternative approaches considered.
This underscores the viability of training models to recognize the note values even in the absence of labeled data through leveraging pseudo-labeling.

\subsection{Input Features}
The proposed framework aims to extract musical information on a singing melody from polyphonic audio. In contrast to the previous methods such as VOCANO\cite{vocano} and MusicYOLO\cite{musicyolo}, which utilize source separation models to isolate vocal melodies from polyphonic audio, this paper adopts an end-to-end approach and takes a polyphonic audio as an input to enable end-to-end transcription of a vocal melody without the need for additional models.

Audio files were converted into waveforms using the madmom library\cite{madmom_python} with a sampling rate of 16,000 Hz. Given that the audio files in the dataset are several minutes long, they were uniformly sliced into the segments of $T$ seconds in length to reduce the model capacity. Subsequently, the audio waveforms are converted into spectrograms that serve as an input for the model. From a variety of spectrogram types such as the short-time Fourier transform (STFT), the Mel transform, and the constant-Q transform, STFT is selected as the input due to its widespread usage across various tasks, including the F0 estimation\cite{jdc} and the note-level singing melody transcription\cite{jong}. The STFT conversion was configured with a hop length of 160 and a window size of 2048. The time frame of the spectrogram was set to 0.01 seconds.

\subsection{Output Representation}

\begin{figure}[!t] 
    \centering
    \includegraphics[width=3.1 in]{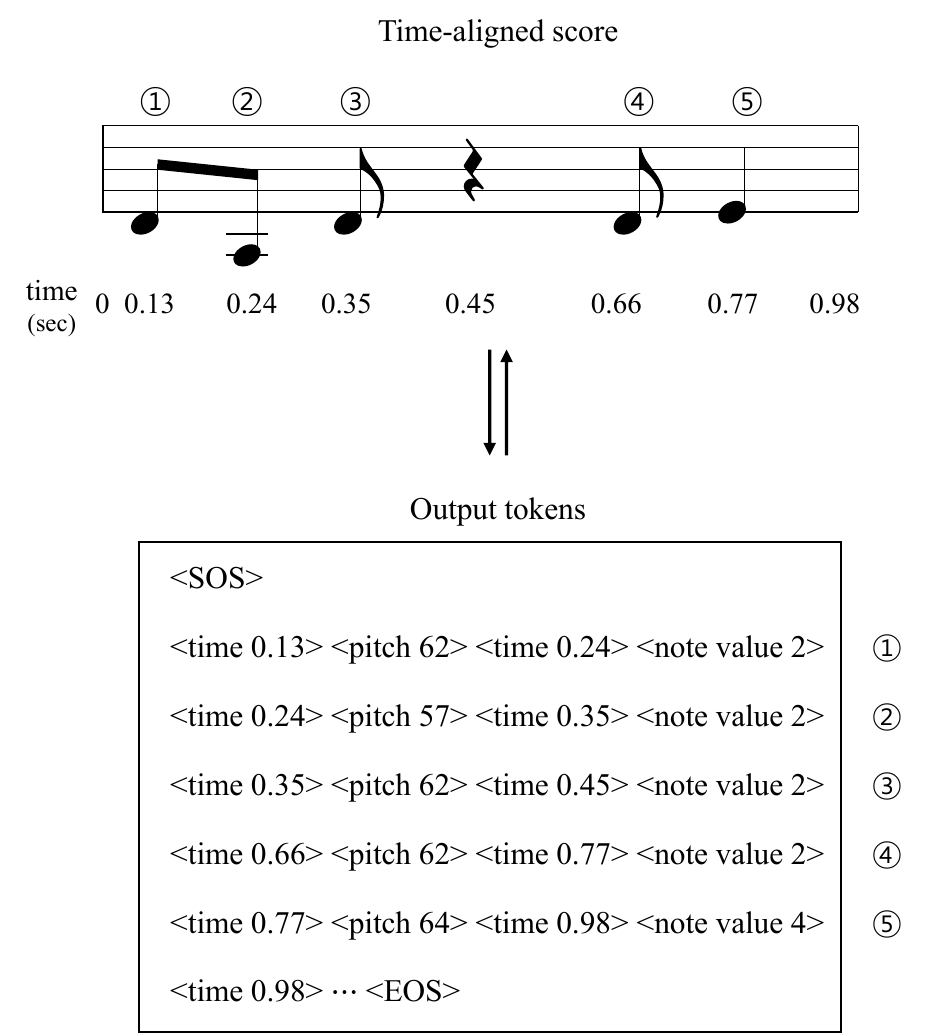}
    \centering
    \caption{Tokenization example. Notes in an audio segment are represented as a token sequence. A note is represented by four tokens: two for time, one for pitch, and one for note value. The start-of-sequence token ($\langle SOS \rangle$) and the end-of-sequence token ($\langle EOS \rangle$) are added to the beginning and the end of the sequence, respectively.}
    \label{fig_token}
\end{figure}

\begin{figure}[!t]
    \centering
    \includegraphics[width=\linewidth]{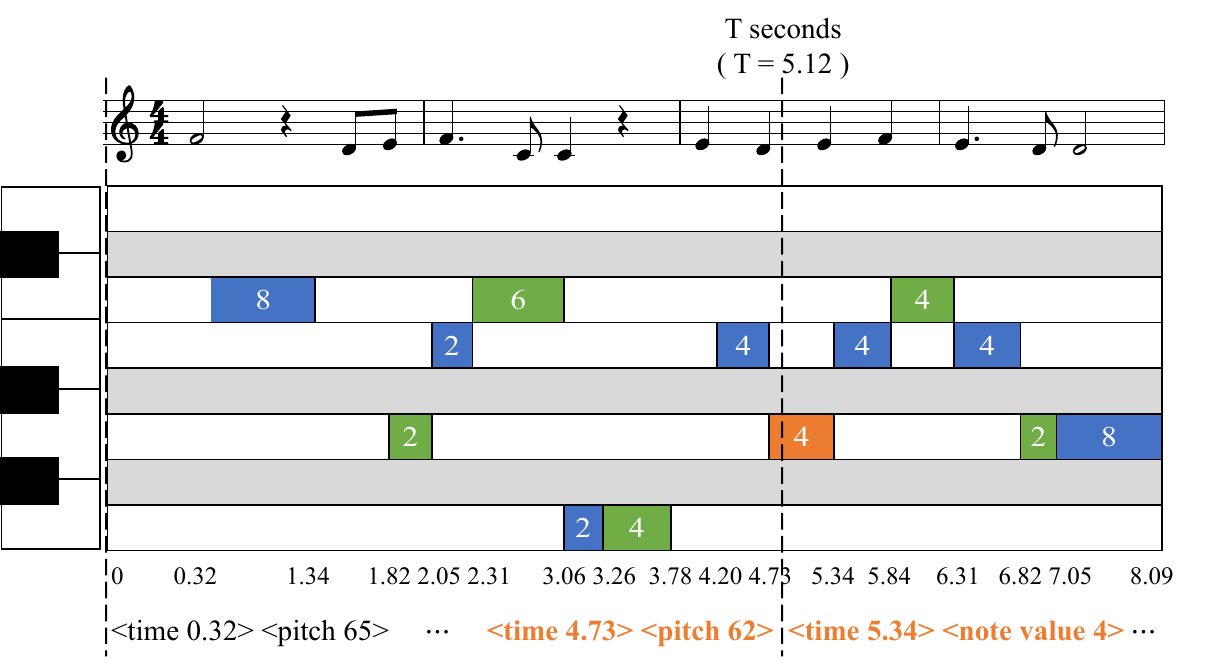}
    \centering
    \caption{Example of segmentation where a note offset exists without its corresponding note onset. The musical score is shown at the top, and the piano-roll representation is shown at the bottom. The numbers in the corresponding piano-roll represent the note values. The note marked in orange is split when the audio is divided into segments of $T$ seconds. The preceding fragment only contains the note onset, while the following fragment only contains the note offset. 
    }
    \label{fig_seg}
\end{figure}

\begin{figure*}[!t]
    \centering
    \includegraphics[width=6in]{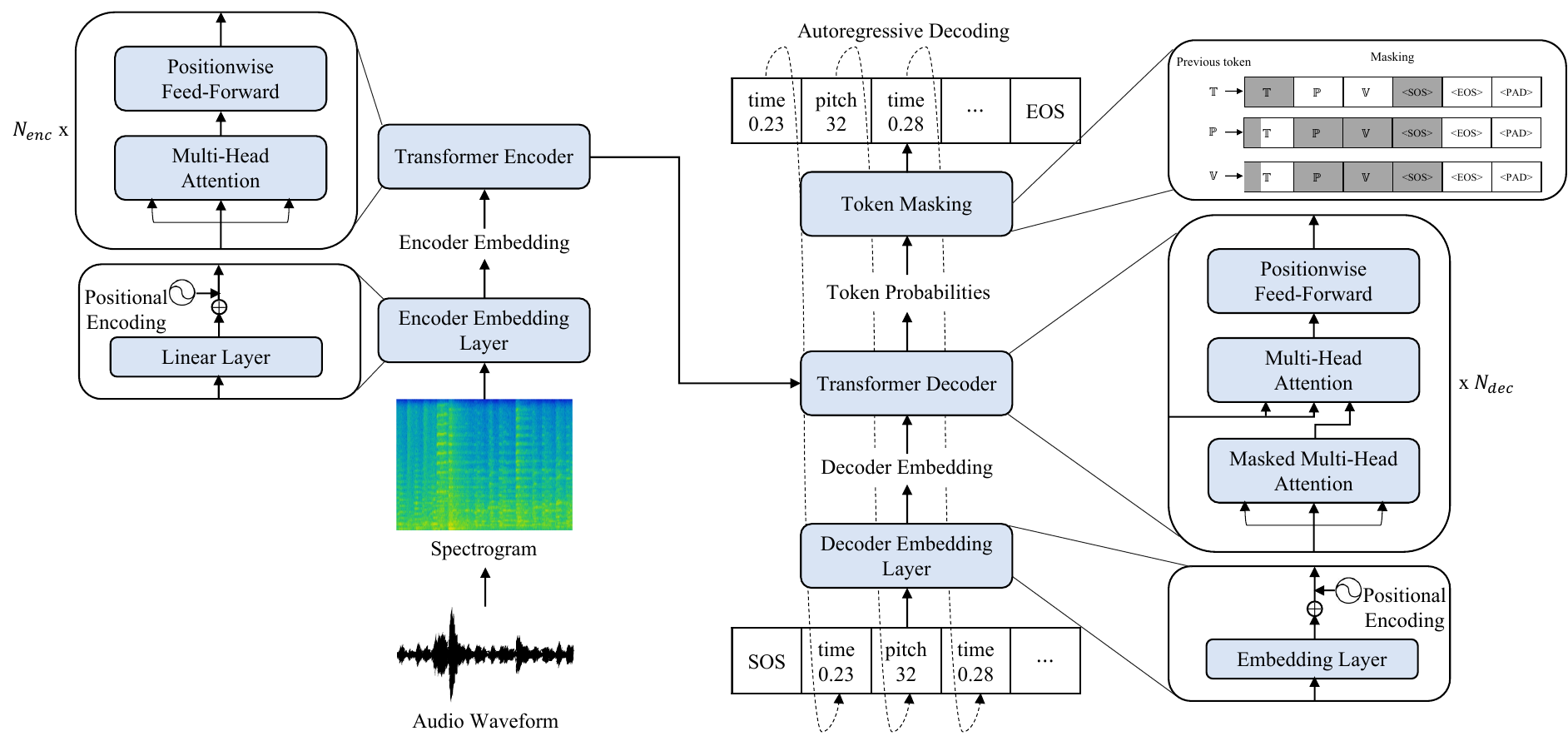}
    \centering
    \caption{Overall architecture of the model within the proposed framework. The audio spectrogram is embedded and the encoder transforms a sequence of audio feature vectors into a sequence of latent representations. The decoder employs the resulting encoder output along with the embedding value of the previous token sequence to compute the probability of each token for predicting a subsequent token at that position. Token masking is applied when computing token probabilities to satisfy the structural requirement of the token sequence. $N_{enc}$ and $N_{dec}$ denote the numbers of Transformer encoder and decoder layers, respectively.}
    \label{fig_model}
\end{figure*}

The output of the proposed framework is a token sequence with the onset time, offset time, pitch, and note value information. The labels for the note-level transcription with the pseudo-labeled note values of $N$ notes are transformed into $y=[\sigma_1, \sigma_2, \dots, \sigma_{4N}]$, where $\sigma_i\in\Sigma$ is the $i^{th}$ token in a token sequence. $\Sigma$ consists of time, pitch, note value, and special tokens: the start-of-sequence token ($\langle SOS \rangle$), the end-of-sequence token ($\langle EOS \rangle$), and the padding token ($\langle PAD \rangle$).
Since a monophonic melody is represented as a single token sequence, $\langle SOS \rangle$, $\langle EOS \rangle$, and $\langle PAD \rangle$ are shared across the sequence, rather than for each time, pitch, and note value. 

The time tokens indicate the positions within the audio segment in terms of absolute time, with a time unit of 0.01 seconds, which corresponds to the time frame of the spectrogram. As a result, for an audio segment of $T$ seconds, there are total of $100T+1$ time tokens. Pitch token ranges from 0 to 127, representing MIDI pitch numbers, and follows the time token. The combination of time and pitch tokens represents the onset time of a note with the specified pitch. 

Note value tokens denote the length of a note in the sixteenth note units, with a range of 1 to $v_{max}$, where $v_{max}$ refers to the maximum note value. Similar to the pitch token, the note value token is located right after the time token. It indicates the offset time of a note with the pre-defined pitch and its note value. The token sequence follows a specific structure where the time token for onset, the pitch token, the time token for offset, and the note value token representing a note are repeated as a subsequence, as defined in (\ref{eq1}).

Special tokens are utilized similarly to those used in natural language processing\cite{nlp_token}.
$\langle SOS \rangle$ is found at the beginning of the label sequence, whereas $\langle EOS \rangle$ is located at the end of the label sequence.
$\langle PAD \rangle$ maintains constant sequence length within a batch, by being inserted after $\langle EOS \rangle$ until the sequence length reaches the maximum length within the batch.

Fig. \ref{fig_token} illustrates transformation of the labels with the onset, offset, pitch, and note value information into a token sequence. Notably, the example excludes $\langle PAD \rangle$ as it represents a single sequence instead of a batch. The onset, offset, and pitch data are provided by the note-level singing transcription datasets, including ST500, while the note value is obtained via pseudo-labeling as described in Section \ref{section_pseudo}. Since only $T$ seconds of the audio is used as an input to the model, there are instances where the audio segment boundary is located between the onset and offset of a note, as shown in Fig. \ref{fig_seg}. In such cases, the information of a single note is split across two segments: the onset time and pitch are included in the preceding segment, while the offset time and note value are included in the subsequent segment.

\subsection{Model Training}

We formulate the transcription task as a sequence-to-sequence problem, with the objective of predicting a sequence of pre-defined tokens from an input sequence of audio features represented as a spectrogram. 

We adopt Transformer\cite{transformer} model from \cite{jong} that already includes task-specific architectural adaptations for transcription tasks. Based on this foundation, we propose further customization for the proposed extended note-level singing transcription task. Specifically, we introduce a token set that represents onset, offset, pitch, and note value information to enable the model to generate time-aligned scores directly from audio input. Sinusoidal positional embeddings and multi-head attentions of Transformer model are expected to improve prediction accuracy by integrating surrounding contextual information. Overall architecture of the model within the proposed framework is depicted in Fig. \ref{fig_model}.

Transformer model is prone to overfitting due to its high model complexity and the large number of parameters, and requires a sufficient amount of data to train the model. We apply random cropping and pitch augmentation during training to enrich the training data and diversify musical contexts.
Random cropping facilitates the learning of diverse time tokens by extracting random audio segments of $T$ seconds from longer audio recordings.
As time tokens are based on absolute time, events at the same time point can be expressed with varying time tokens depending on the time frame from which the audio segment starts. 
For the pitch augmentation, pitch is shifted by a randomly selected semitone within a pre-defined range of $-$6 to 6. The input audio is transformed by shifting the pitch of the waveform using the pysndfx library\footnote{https://github.com/carlthome/python-audio-effects}, and the output sequence is adjusted accordingly through modifying the pitch tokens. Pitch augmentation is expected to enable the learning of various pitch tokens. 

The contributions of this paper lie in integrating note value recognition into an end-to-end transcription framework to address the challenges of time-aligned score generation, as well as acheiving performance improvements solely through the definition of the new task and framework, without requiring major modifications to the existing model architecture.

\subsection{Inference}
For inference, the model requires only the audio as an input. The audio is sliced into short segments of $T$ seconds and the recognition results from multiple segments are combined to produce the final output.
The spectrogram corresponding to $T$-second is passed through the model encoder to produce latent representations.
The model decoder predicts a next token in an autoregressive manner using the latent representations of audio from the encoder and the previous token sequence. When the previous token sequence is empty, $\langle SOS \rangle$ is added to the sequence. This prediction process continues until the decoder predicts $\langle EOS \rangle$ or the decoder output reaches a pre-determined maximum output length.

Fundamentally, as defined in (\ref{eq2}) and (\ref{eq3}), the token sequence is determined by selecting the next token with the highest probability, regardless of the token category. However, from the definition of the token set, a time token must be followed by either a pitch or note value token, and a pitch or note value token must be followed by a time token. To ensure adherence to this structure, the output probabilities of the model are masked, following the approach described in \cite{jong}. Specifically, if the previous token is a time token, subsequent time tokens are masked, and if the previous token is a pitch or note value token, both pitch and note value tokens are masked. Time tokens corresponding to time frames earlier than the previously selected time token are also masked to ensure monophonic melody transcription, as illustrated in Fig. \ref{fig_model}.

Despite these masking strategies, invalid token sequences such as $(\hat{t}_{i}, p_i, \hat{t}_{i+1}, p_{i+1})$ or $(\doublehat{t}_{i}, v_i, \doublehat{t}_{i+1}, v_{i+1})$ may occur during inference. These are ignored during the transformation of token sequences into note sequences to ensure that only valid token sequences like $(\hat{t}_{i}, p_i, \doublehat{t}_{i}, v_{i})$ are considered. Fortunately, such cases were infrequent when utilizing a well-trained model. In the evaluation of 100 songs from the ST500 test set, 81 invalid cases were observed, accounting for only 0.26\% of the 31,175 valid sequences.

We employ an overlapping decoding method\cite{jong} that slices audio segments with an overlap and leverages the recognition results from the previous segment as decoder input to predict the result of the subsequent segment. This approach mitigates inconsistencies in the recognition results across continuous audio segments while enhancing the recognition performance of note values by reducing the number of instances where the note offset exists without the corresponding note onset.

\begin{algorithm}[t]
\caption{Rest and barline estimation}
\label{alg_score}
\begin{algorithmic}[1]

\Require $y = \left[(\hat{t}_{1}, p_1, \doublehat{t}_{1}, v_1), \dots, (\hat{t}_{N}, p_N, \doublehat{t}_{N}, v_N) \right]$
\State $\textit{score} \leftarrow [\text{CreateNote}(p_1, 0, v_1)]$
\State $\textit{beatPosition} \leftarrow v_1$
\For{$i = 2$ to $N$}
    \State $\Delta t \leftarrow \hat{t}_{i} - \doublehat{t}_{i-1}$
    \If{$\Delta t > 0$}
        \State $d_{\text{avg}} \leftarrow \frac{1}{2}((\doublehat{t}_{i-1} - \hat{t}_{i-1}) / {v_{i-1}} + (\doublehat{t}_{i} - \hat{t}_{i}) / {v_{i}})$
        \State $r \leftarrow \text{round}\left(\Delta t / d_{\text{avg}}\right)$
        \If{$r > 0$}
            \State $\textit{score}.\text{append}(\text{CreateRest}(beatPosition, r))$
            \State $\textit{beatPosition} \leftarrow \textit{beatPosition} + r$
        \EndIf
    \EndIf
    \State $\textit{score}.\text{append}(\text{CreateNote}(p_i, beatPosition, v_i))$
    \State $\textit{beatPosition} \leftarrow \textit{beatPosition} + v_i$
\EndFor

\For{$b = 1$ to $\left\lfloor \textit{beatPosition} / 16 \right\rfloor$}
    \State $\textit{score}.\text{append}(\text{CreateBarline(16 $\times$ b)})$
\EndFor
\If{$\textit{beatPosition} \mod 16 > 0$} \label{lstart}
    \State $r \leftarrow 16 - (\textit{beatPosition} \mod 16)$
    \State $\textit{score}.\text{append}(\text{CreateRest}(beatPosition, r))$
    \State $\textit{score}.\text{append}(\text{CreateBarline}(beatPosition + r)$
\EndIf \label{lend}

\Return $\textit{score}$

\end{algorithmic}
\end{algorithm}

{\subsection{Score Visualization}}
The model generates a token sequence representing notes with pitches from the music audio. Since rests are not explicitly recognized, a heuristic algorithm predicts rests and inserts barlines to visualize the transcription results in terms of a musical score. The detailed procedure for constructing a time-aligned musical score from a token sequence is described in Algorithm \ref{alg_score}, which involves the following functions:
\begin{itemize}
    \item \textbf{CreateNote($p$, $b$, $v$)}: Creates a note symbol with pitch $p$, starting at beat position $b$, and lasting for note value $v$.
    \item \textbf{CreateRest($b$, $r$)}: Creates a rest symbol starting at beat position $b$ with note value $r$.
    \item \textbf{CreateBarline($b$)}: Adds a barline at beat position $b$ to mark the end of a measure.
\end{itemize}

When the time gap between the offset of one note and the onset of the next note ($\Delta t$) is greater than zero, the note value of the rest ($r$) is determined by dividing $\Delta t$ by the estimated duration of a sixteenth note ($d_\text{avg}$) and rounding to the nearest integer. The value of $d_\text{avg}$ is calculated as the average duration of the sixteenth notes from the preceding and succeeding notes. If $r$ is greater than 0, a rest with the estimated note value $r$ is appended to the score between the two notes. $beatPosition$ tracks the cumulative beat position in the sixteenth note units and is updated by adding the note value of each newly appended note or rest, thereby marking the start position of the next symbol. Assuming a fixed time signature of 4/4, barlines are added every 16 note values to signify the end of a measure. This approach inherently assumes that the first measure starts with the first predicted note, without considering the cases where a rest precedes the first note. At the end of the sequence, any remaining beats are filled with a final rest and a closing barline to complete the score (Lines \ref{lstart}--\ref{lend}).

Algorithm \ref{alg_score} generates a sequence of notes, rests, and barlines under the assumption that the predictions of the model reflect the rhythmic structure of the audio. The resulting score provides a time-aligned representation with predicted onset and offset times, serving as a foundation for qualitative evaluation, discussed in Section \ref{section_qual_eval}.

\section{Experiments}
\subsection{Datasets}
Several datasets are available for the note-level singing melody transcription, including ST500\cite{st500}, DALI\cite{dali}, TONAS\cite{tonas}, and SSVD\cite{musicyolo}. The audio files in TONAS and SSVD contain only human voice, which differentiates them from pop songs that include instrumental accompaniment. DALI contains a large amount of 5,358 songs with accompaniment, but it contains errors in timing or pitch octave due to automatic alignment of labels. ST500, consisting of 500 pop songs, was labeled through iterative process of labeling, verification, and post-processing. It is considered as a valuable dataset for training a feasible AST model\cite{st500}. 

For this study, ST500 was selected for pseudo-labeling. 474 songs, which can be downloaded from YouTube, were used in the experiments. The data were split into three sets: songs 1 to 350 for training, 351 to 400 for validation, and 401 to 500 for testing. This splitting strategy aligns with the previous studies\cite{st500, jong, musicyolo} and allows comparisons under the same conditions. 
The test set was utilized to evaluate the note-level transcription performance of each model, in terms of onset, offset, and pitch while excluding note values. All test audio files were available for download, enabling consistent comparisons with the previous methods.

Few datasets provide the note value labels, along with their corresponding audio files. Notation-level transcription\cite{notation1, notation2, notation3} that predicts pitches and note values employed MIDI-synthesized audio due to the absence of audio files.
Datasets with onset, offset, pitch, and note value information include RWC Music Database: Popular Music\cite{rwc} and HSD\cite{hsd}. RWC Music Database: Popular Music is not annotated for transcription and its timing and pitch octave accuracy may be unsatisfactory for model training.
HSD includes lyrics and hierarchical structures in addition to the onset, offset, pitch, and note value. While the manually calibrated labels of HSD was quitely reliable, it was utilized for evaluation purposes only in our experiments due to its limited availability.

Note values are expressed as floating point units based on the quarter note in the HSD dataset but as natural numbers in units of the sixteenth note in this paper. To unify the note value representation, the note value labels in HSD were converted to the units of the sixteenth note.
Additionally, discrepancies were observed between some labels in HSD and the corresponding audio, particularly with respect to pitch octave and timing. To address these discrepancies, JDC\cite{jdc} was utilized to detect F0 of the audio. The labels were then adjusted by shifting the pitch octave from $-$3 to 3 with an interval of one octave and the timing from $-$3 seconds to 3 seconds with an interval of 10 msec. The label with the highest overall F0 accuracy\cite{f0_eval} was selected, and the labels with an overall F0 accuracy of 0.55 or lower, even after adjustments of the pitch octave and timing, were excluded from the test set. Note value recognition performance was evaluated using the resulting HSD dataset, which consists of 55 songs with the modified note value representations, pitch octave, and timing.

\subsection{Training Details}
\begin{table}
\centering
\caption{Hyperparameters of the proposed model}
\label{model_config}
\begin{tabular}{l|ll}
\hline
Hyperparameter                     & Encoder & Decoder \\ \hline
Number of layers                   & 12      & 8       \\
Dimension of embedding             & 512     & 512     \\
Number of self-attention heads     & 8       & 8       \\
Dimension of query, key, value     & 512     & 512     \\
Dropout probability                & 0.1     & 0.1     \\
Maximum length of  output sequence & -       & 512     \\ \hline
\end{tabular}
\end{table}
The hyperparameters of the model considered in the experiments are listed in Table \ref{model_config}. For model optimization, the cross-entropy loss function and the Adam optimizer were employed. The cyclic learning rate scheduler\cite{cycliclr} that adjusts a learning rate with a maximum value of 0.001 was utilized. Early stopping was implemented if the validation loss did not decrease for 10 consecutive epochs.

The audio signal was divided into the segments of 5.12 seconds. The maximum note value in the sixteenth note units, $v_{max}$, was set to 32. 
The token set has 676 tokens, comprising 513 time tokens, 128 pitch tokens, 32 note value tokens, as well as three special tokens: $\langle SOS \rangle$, $\langle EOS \rangle$, and $\langle PAD \rangle$.

For inference, both the batch method and overlapping decoding method were implemented to compare their performances.
In the batch method, the batch size was set to 8.
In the overlapping decoding method, the hop size was set to 2.56 seconds, and the last 1.28 seconds of each segment were discarded, as in the previous research\cite{jong}.

\subsection{Evaluation Metrics}
\label{section_metric}
Performance of transcribing the onset, offset, and pitch without considering the predicted note value can be evaluated using the standard note-level transcription evaluation metrics, commonly employed in the existing note-level transcription studies\cite{ctc_note, musicyolo, jong, av-svt}.
The note-level transcription performance is evaluated in four ways, depending on the elements considered for assessing the correctness of the estimated notes: onset only, offset only, onset and pitch, and overall note-level transcription incorporating onset, offset, and pitch. 

Correct detection of the onset, offset, and pitch in these metrics indicates that the difference between the label and the predicted value is below a threshold, consistent with the mir\_eval\cite{mir_eval} library and the previous research.
The threshold values for the onset, offset, and pitch were 50 msec, 50 msec or 20\% of the time between the ground-truth onset and offset, whichever is greater, and 0.5 semitones, respectively. That is: 
\begin{gather}
\vert \hat{t}_{i}-\bar{t}_{i}\vert\leq 50 \,\text{msec},
\label{co} \\
\vert \doublehat{t}_{i} - \dbar{t}_{i}\vert \leq \max\left(50 \,\text{msec}, \, 0.2 \times (\dbar{t}_{i}-\bar{t}_{i})\right),
\label{coff}\\
\vert {p}_{i}-\bar{p}_i\vert\leq 50 \,\text{cents},
\label{cp}
\end{gather}
where $\hat{t}_{i}$, $\doublehat{t}_{i}$, and ${p}_i$ denote the predicted onset time, offset time, and pitch of the $i^{th}$ note in a note sequence, respectively, while $\bar{t}_{i}$, $\dbar{t}_{i}$, and $\bar{p}_i$ represent the corresponding ground-truth values, respectively.
The evaluation metrics include precision, recall, and F1 score, calculated for each of the four evaluation aspects using the mir\_eval library.
The formulas for calculating precision ($P$), recall ($R$), and F1 score ($F$) are as follows\cite{eval_prf}:
\begin{gather}
P = \frac{TP}{TP+FP} \text{ ,} \\
R = \frac{TP}{TP+FN} \text{ ,} \\
F = \frac{2 \times P \times R}{P+R} \text{ ,}
\end{gather}
where $TP$, $FP$, and $FN$ denote the numbers of correctly recognized notes, incorrectly recognized notes, and missed notes, respectively.

Currently, to the best of our knowledge, no existing metric simultaneously evaluates the transcription performance of the onset times, offset times, pitches, and note values, since the previous research recognizing the note values did not consider the timing information.
Therefore, we propose new metrics that consider the note values in addition to the note-level transcription evaluation metrics.
When the precision, recall, and F1 scores were calculated, the correctness of the predicted notes is determined by three criteria: (1) when the onset, offset, pitch, and note value are all correctly predicted, (2) when the onset, offset, and note value are correctly predicted, and (3) when the onset, pitch, and note value are correctly predicted. The correct recognition of the onset, offset, and pitch was determined by using the same methods and thresholds as described in (\ref{co}), (\ref{coff}), and (\ref{cp}).

Correctness of the predicted note value was decided by $\vert v_i-\bar{v}_i\vert$, where $v_i$ and $\bar{v}_i$ represent the predicted note value and the ground-truth note value of the $i^{th}$ note in a note sequence, respectively.
If the difference is less than the threshold, we consider the predicted note value to be correct. 

\begin{figure}[!t]
    \centering
    \includegraphics[width=2.5in]{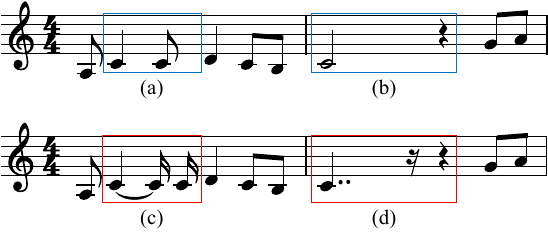}
    \centering
    \caption{Incorrect note values resulted from beat quantization using incorrect beat tracking results. The correct score is shown at the top, and the score from beat quantization is shown at the bottom. Notes with incorrect note values are marked in red, and the corresponding correct notes are marked in blue.}
    \label{fig_bq}
\end{figure}

One of the main problems with the beat quantization is the discrepancy between the predicted and actual note values due to the irregularities in the recognized beats.
Fig. \ref{fig_bq} shows examples of incorrect note values resulting from the beat quantization. In the first bar, two consecutive notes in (c) were quantized to longer and shorter note values, respectively, compared to the note values of the ground-truth notes in (a). The opposite case is shown in the second bar. The preceding note in (d) was quantized to shorter note value than the ground-truth note in (b), while the subsequent rest in (d) was quantized to longer note value than the ground-truth rest in (b). 

Since even a difference of just a sixteenth note may appear unnatural due to the presence of numerous ties or notes with double dots, we set the note value threshold to zero, implying that a predicted note value is considered correct only when it exactly matches the label. That is, the criterion for correctly predicted note value is:
\begin{gather}
    \vert v_i-\bar{v}_i\vert = 0.
\end{gather}

The proposed metrics are inherently constrained by the note-level transcription performance and incorrectly transcribed notes in terms of onset, offset, and pitch cannot provide a meaningful basis for evaluating note values. To address this limitation, accuracy and Mean Squared Error (MSE) are introduced to assess the note value estimation performance independently of errors in onset, offset, or pitch detection. Accuracy measures the proportion of the correctly recognized note values and MSE quantifies the degree of difference between the predicted note values and the ground-truth note values. These evaluations are applied exclusively to correctly recognized notes, considered under two conditions: CN1 where the onset and offset were correctly recognized, and CN2 where the onset, offset, and pitch were correctly recognized.

Accuracy and MSE should be interpreted alongside the note-level transcription performance metrics discussed earlier, as they are meaningful only when comparing models with similar note-level transcription performance. While note-level metrics assess the overall transcription performance, MSE and accuracy offer a fine-grained evaluation of the note value prediction performance. This combined evaluation framework provides a more comprehensive understanding of the prediction results.

To compare with the existing research on the note value recognition without the time-aligned information, we also evaluated the transcription performance in the form of a lead sheet that contains only the pitches and note values.
Research methods that recognize the note values include rhythm quantization\cite{nakamura_rhythm}, notation-level transcription\cite{notation1, notation2, notation3}, and audio-to-symbolic singing melody transcription\cite{weakly_attention, beat_attention, CRNN_HSMM}. However, the metrics used to evaluate the recognition performance in each study were not standardized and varied widely.

In this paper, we adopted the error rate metrics used in the previous studies such as \cite{weakly_attention} and \cite{beat_attention}. These metrics assess the estimated pitches and note values of a monophonic singing melody.
The error rate ($ER$) is defined as follows:

\begin{equation}
\label{eq_er}
ER = \frac{N_S + N_D + N_I} {N} \times 100\,(\%),
\end{equation}
where $N_S$, $N_D$, and $N_I$ represent the minimum numbers of substitutions, deletions, and insertions required to transform the predicted result into the ground-truth label, respectively. The numerator represents the Levenshtein distance\cite{leven} as the sum of $N_S$, $N_D$, and $N_I$. The denominator refers to the number of notes present in the ground-truth label as $N$.
We used three error rates as evaluation metrics: (1) pitch error rate (PER) that considers only pitch, (2) value error rate (VER) that considers only note value, and (3) note error rate (NER) that considers both pitch and note value.

\subsection{Models Compared}
To assess the note-level transcription performance, we compared our proposed model with the existing models such as AV-SVT\cite{av-svt, altamt}, CE+CTC\cite{ctc_note}, note-level Transformer\cite{jong}, and MusicYOLO-I \cite{musicyolo}. These models were trained using the same data split for ST500 to enable fair comparisons. 
AV-SVT is a singing voice transcription system based on self-supervised models from the speech domain.
CE+CTC, trained with both CTC loss and cross-entropy loss, utilizes Spleeter\cite{spleeter} for input pre-processing.
Note-level Transformer is a Transformer-based model using ST500 and filtered DALI data. MusicYOLO-I used vocal audio extracted by Spleeter as input and performed transcription through YOLOX-based\cite{yolox} object detection with pitch labeling.

Unfortunately, there are few end-to-end models capable of recognizing the onset and offset times, pitch, and note value. Published audio-to-symbolic singing melody transcription models that can directly estimate the pitches and note values from audio are also limited.
Accordingly, we compared the note values obtained by quantizing the note-level transcription results and the note values recognized directly from the audio using the proposed model.
Note-level Transformer\cite{jong} was used for the note-level transcription and the transcription results were quantized in two ways: grid quantization using the beat tracking results by madmom and the rhythm quantization using commercial software MuseScore\cite{musescore}. To compare performance in the symbolic domain, we utilized AMT software AnthemScore\cite{anthemscore} to consider pitches and note values while excluding the onset and offset times.

\section{Results}

\begin{table}[]
\centering
\caption{Model configurations for the preliminary experiment and ablation study}
\label{table_config}

\begin{tabular}{c|ccc|cccc}
\hline
                        & \multicolumn{3}{c|}{Preliminary Experiments} & \multicolumn{4}{c}{Ablation Studies}                        \\ \cline{2-8} 
\multirow{-2}{*}{model} & SR (Hz)          & OD          & DALI        & Input          & $\mathbb{T}$ & $\mathbb{P}$ & $\mathbb{V}$ \\ \hline
E1                      & 16,000            & \textbf{-}  & -           & Raw            & O            & O            & O            \\
\rowcolor{gray!20} 
E2 (\textbf{T3MS})                   & 16,000            & O           & -           & Raw            & O            & O            & O            \\
E3                      & 16,000            & O           & \textbf{O}  & Raw            & O            & O            & O            \\
E4                      & \textbf{44,100}   & O           & -           & Raw            & O            & O            & O            \\ \hline
Note-level T3MS         & 16,000            & O           & -           & Raw            & O            & O            & \textbf{X}   \\
Symbolic T3MS           & 16,000            & O           & -           & Raw            & \textbf{X}   & O            & O            \\
T3MS \textbf{+ GQ}               & 16,000            & O           & -           & Raw            & O            & O            & O            \\
Vocal T3MS              & 16,000            & O           & -           & \textbf{Vocal} & O            & O            & O            \\ \hline
\end{tabular}
\end{table}

\begin{table*}
\centering
\caption{Note-level transcription performance of the preliminary experiments and ablation studies on the ST500 test set}
\label{table_ours_note}
\begin{tabular}{c|ccc|ccc|ccc|ccc}
\hline
                        & \multicolumn{3}{c|}{Onset}                                                         & \multicolumn{3}{c|}{Offset}                                                        & \multicolumn{3}{c|}{Onset + Pitch}                                                 & \multicolumn{3}{c}{\begin{tabular}[c]{@{}c@{}}Note-level\\ (Onset+Offset+Pitch)\end{tabular}} \\
\multirow{-3}{*}{model} & P                         & R                         & F                          & P                         & R                         & F                          & P                         & R                         & F                          & P                             & R                             & F                             \\ \hline
E1                      & 0.810           & 0.775                     & 0.790                      & 0.764            & 0.729                     & 0.744                      & 0.771                     & 0.739                     & 0.753                      & 0.606                         & 0.581                         & 0.592                         \\
\rowcolor{gray!20}
E2 (T3MS)                 & 0.808                     & 0.808                     & 0.806             & 0.761                     & 0.760            & 0.759             & 0.772            & 0.772                     & 0.771                      & 0.610                & 0.611                & 0.610                \\
E3                      & 0.804                     & 0.811            & 0.806             & 0.748                     & 0.754                     & 0.750                      & 0.772            & 0.779            & 0.775             & 0.605                         & 0.610                         & 0.607                         \\
E4                      & \textbf{0.821}                     & \textbf{0.822}                     & \textbf{0.820}                      & \textbf{0.768}                     & \textbf{0.769}                     & \textbf{0.767}                      & \textbf{0.783}                     & \textbf{0.784}                     & \textbf{0.783}                      & \textbf{0.624}                         & \textbf{0.625}                         & \textbf{0.624}                         \\ \hline
Note-level T3MS & {0.657} & {0.800} & {0.718} & {0.621} & {0.756} & {0.678} & {0.627} & {0.764} & {0.685} & {0.491} & {0.598} & {0.537} \\
Vocal T3MS      & {0.671} & {0.685} & {0.677} & {0.717} & {0.732} & {0.723} & {0.637} & {0.650} & {0.643} & {0.489} & {0.500} & {0.494} \\ \hline
\end{tabular}
\end{table*}

\subsection{Preliminary Experiment Results}
Table \ref{table_config} outlines the configurations of models, named E1, E2, E3, and E4, used in the preliminary experiments. These experiments attepted to analyze three key aspects of the proposed model: the overlapping decoding method (OD), the use of the filtered DALI data, and the sampling rate (SR), as shown in Table \ref{table_config}. For each configuration, all other settings remained identical except for the factor under comparison.

First, E1, E2, and E3 were compared to investigate the effects the decoding methods and the impact of the filtered DALI data. The decoding methods involve the batch method and the overlapping decoding method. To assess the effect of increasing the amount of training data using the filtered DALI data, data cleansing and filtering were performed following \cite{jong}.

Table \ref{table_ours_note} provides a comparative analysis of the note-level singing melody transcription performances on the ST500 test set. 
The results indicate that the overlapping decoding method outperforms the batch method, owing to the enhanced consistency between the consecutive segments achieved by utilizing the recognition results of the previous segment when recognizing the subsequent segment. 

\begin{table*}
\centering
\caption{Note value recognition performance from the preliminary experiments and ablation studies on the HSD test set}
\label{table_ours_nl}
\begin{tabular}{c|ccc|ccccc|ccccc}
\hline
                        & \multicolumn{3}{c|}{Onset + Pitch + Note Value}  & \multicolumn{5}{c|}{Onset + Offset + Note Value {(CN1)}}                                   & \multicolumn{5}{c}{Onset + Offset + Pitch + Note Value {(CN2)}}                            \\
\multirow{-2}{*}{model} & P              & R              & F              & P              & R              & F              & acc            & MSE            & P              & R              & F              & acc            & MSE            \\ \hline
E1                      & 0.391          & 0.372          & 0.381          & 0.361          & 0.342          & 0.351          & 0.678          & 0.490          & 0.341          & 0.323          & 0.331          & 0.532          & 0.822          \\
\rowcolor{gray!20}
E2 (T3MS)               & 0.403          & 0.398          & 0.400          & 0.373          & 0.367          & 0.369          & 0.688          & 0.470          & 0.354          & 0.349          & 0.351          & 0.546          & 0.796          \\
E3                      & \textbf{0.428} & \textbf{0.425} & \textbf{0.426} & \textbf{0.393} & \textbf{0.390} & \textbf{0.391} & \textbf{0.723} & \textbf{0.419} & \textbf{0.377} & \textbf{0.374} & \textbf{0.375} & \textbf{0.577} & \textbf{0.757} \\
E4                      & 0.390          & 0.385          & 0.387          & 0.357          & 0.352          & 0.355          & 0.667          & 0.511          & 0.339          & 0.334          & 0.336          & 0.529          & 0.835          \\ \hline
{T3MS + GQ}                       & {0.390} & {0.385} & {0.387} & {0.356} & {0.351} & {0.353} & {0.665} & {0.495} & {0.339} & {0.334} & {0.336} & {0.531} & {0.823} \\
{Vocal T3MS}                      & {0.345} & {0.343} & {0.344} & {0.311} & {0.309} & {0.309} & {0.689} & {0.445} & {0.296} & {0.294} & {0.295} & {0.555} & {0.797} \\ \hline
\end{tabular}
\end{table*}

\begin{table}[]
\centering
\caption{Symbolic error rates from the ablation studies on the HSD test set}
\label{table_ours_sym}
\begin{tabular}{c|ccc|ccc}
\hline
{}                        & \multicolumn{3}{c|}{{note only}}                                                              & \multicolumn{3}{c}{{note and rest}}                                                           \\
\multirow{-2}{*}{{model}} & {NER}           & {PER}           & {VER}           & {NER}           & {PER}           & {VER}           \\ \hline
{T3MS}                    & {\textbf{64.4}} & {\textbf{11.5}} & {46.4}          & {\textbf{78.6}} & {\textbf{28.8}} & {\textbf{60.0}} \\
{T3MS + GQ}               & {65.9}          & {\textbf{11.5}} & {47.5}          & {79.8}          & {\textbf{28.8}} & {61.1}          \\
{Vocal T3MS}              & {65.6}          & {12.8}          & {\textbf{46.1}} & {83.9}          & {35.2}          & {64.8}          \\
{Symbolic T3MS}           & {510.5}         & {281.7}         & {259.1}         & {527.2}         & {313.3}         & {291.5}         \\ \hline
\end{tabular}
\end{table}

Table \ref{table_ours_nl} compares the note value recognition performance on the HSD test set using the metrics proposed in Section \ref{section_metric}.
The results indicate superior performances across all the metrics when employing the overlapping decoding method. This performance enhancement is attributed to the ability to address the issues arising from splitting the onset and offset of a note into different segments. 

Although the inclusion of the filtered DALI data improves the note value recognition performance, it does not improve the note-level transcription performance. In contrast, the overlapping decoding method enhances both aspects of performance. Based on these findings, the configuration of E2 was selected for the subsequent experiments.

Second, the selected configuration of E2 was analyzed in terms of sampling rate. Previous note-level transcription studies have utilized various sampling rates without a standardized choice. To address this, preliminary experiments compared 16,000 Hz (E2) and 44,100 Hz (E4).

Table \ref{table_ours_note} and Table \ref{table_ours_nl} show that E2 outperformed E4 in recognizing the note values. Although the higher sampling rate of E4 allows finer time resolution for improved note-level transcription, it may also introduce unnecessary noise that adversely affects the note value prediction. Consequently, 16,000 Hz was selected as the optimal sampling rate due to its superior note value prediction performance and lower computational complexity. The proposed model with the selected configuration of E2 is referred to as T3MS (\textbf{T}ransformer for \textbf{T}ranscribing \textbf{T}ime-aligned \textbf{M}usical \textbf{S}core).

\subsection{Ablation Study}
We conducted ablation studies to assess the impact of different model components and input types on the extended note-level transcription task. 
Table \ref{table_config} summarizes the model configurations used in the ablation study, which share the same model architecture as T3MS: (1) note-level T3MS, which performs conventional note-level transcription without recognizing the note values, (2) symbolic T3MS, which recognizes only the pitch and note value in the symbolic domain without temporal information, (3) T3MS + GQ that estimates the note values by applying the grid quantization to the recognized onset and offset times instead of using the note values predicted by T3MS, and (4) vocal T3MS that uses vocal-separated audio as input. The results of the ablation study are presented in Table \ref{table_ours_note}, Table \ref{table_ours_nl}, and Table \ref{table_ours_sym}. The symbolic T3MS was evaluated solely based on symbolic error rates, as it cannot recognize onset and offset times, whereas the note-level T3MS was assessed exclusively for the note-level transcription performance due to its inability to estimate note values.

The results demonstrate that T3MS outperformed all the other models across most metrics. The vocal T3MS performed worse than T3MS in most metrics, primarily due to noise introduced during the vocal separation process. However, its higher accuracy and lower MSE based on the correctly recognized notes can be attributed to enhanced note value recognition performance provided by the accurately extracted vocal part. The symbolic T3MS performed significantly worse across all the error metrics due to its lack of time information and inability to employ overlapping decoding. Moreover, CTC loss, which is better suited for symbolic transcription tasks, was not utilized during its training.

The superior performance of T3MS over the note-level T3MS in note-level transcription and the symbolic T3MS in symbolic transcription underscores the effectiveness of jointly estimating time and note value in T3MS. By leveraging the mutual reinforcement between these tasks, T3MS outperformed the models designed for individual tasks. This was further evidenced by its comparison with T3MS+GQ, where T3MS achieved better results than T3MS+GQ across all metrics. T3MS+GQ estimates note values from the predicted onset and offset times, ignoring the predicted note values from T3MS.
These findings suggest that the remarkable performance of T3MS stems not only from its note-level transcription performance but also from its enhanced note value prediction, surpassing the results obtained through grid quantization.

\subsection{Comparison of Note-Level Transcription Performance}
\label{section_hsd}

\begin{table*}
\centering
\caption{Comparison results for note-level transcription performances evaluated on either the ST500 or HSD test set}
\label{table_note}
\begin{tabular}{c|c|lll|lll|lll|lll}
\hline
                                              &                         & \multicolumn{3}{c|}{Onset}                                                 & \multicolumn{3}{c|}{Offset}                                            & \multicolumn{3}{c|}{Onset + Pitch}                                         & \multicolumn{3}{c}{\begin{tabular}[c]{@{}c@{}}Note-level\\ (Onset+Offset+Pitch)\end{tabular}} \\
\multirow{-3}{*}{dataset}                     & \multirow{-3}{*}{model} & \multicolumn{1}{c}{P} & \multicolumn{1}{c}{R} & \multicolumn{1}{c|}{F}     & \multicolumn{1}{c}{P} & \multicolumn{1}{c}{R} & \multicolumn{1}{c|}{F} & \multicolumn{1}{c}{P} & \multicolumn{1}{c}{R} & \multicolumn{1}{c|}{F}     & \multicolumn{1}{c}{P}        & \multicolumn{1}{c}{R}        & \multicolumn{1}{c}{F}           \\ \hline
                                              & AE-SVT                  & \multicolumn{1}{c}{-} & \multicolumn{1}{c}{-} & \multicolumn{1}{c|}{0.781} & \multicolumn{1}{c}{-} & \multicolumn{1}{c}{-} & \multicolumn{1}{c|}{-} & \multicolumn{1}{c}{-} & \multicolumn{1}{c}{-} & \multicolumn{1}{c|}{0.700} & \multicolumn{1}{c}{-}        & \multicolumn{1}{c}{-}        & \multicolumn{1}{c}{0.528}       \\
                                              & CE+CTC                  & \multicolumn{1}{c}{-} & \multicolumn{1}{c}{-} & \multicolumn{1}{c|}{0.796} & \multicolumn{1}{c}{-} & \multicolumn{1}{c}{-} & \multicolumn{1}{c|}{-} & \multicolumn{1}{c}{-} & \multicolumn{1}{c}{-} & \multicolumn{1}{c|}{0.744} & \multicolumn{1}{c}{-}        & \multicolumn{1}{c}{-}        & \multicolumn{1}{c}{0.574}       \\
                                              & Note-level Transformer  & 0.785                 & 0.791                 & 0.787                      & 0.747                 & 0.753                 & 0.749                  & 0.754                 & 0.761                 & 0.757                      & 0.589                        & 0.595                        & 0.591                           \\
                                              & MusicYOLO-I             & \textbf{0.819}        & 0.750                 & 0.782                      & \textbf{0.788}        & 0.720                 & 0.751                  & 0.747                 & 0.686                 & 0.714                      & \textbf{0.612}               & 0.563                        & 0.586                           \\
\multirow{-5}{*}{ST500}                       & T3MS                    & 0.808                 & \textbf{0.808}        & \textbf{0.806}             & 0.761                 & \textbf{0.760}        & \textbf{0.759}         & \textbf{0.772}        & \textbf{0.772}        & \textbf{0.771}             & 0.610                        & \textbf{0.611}               & \textbf{0.610}                  \\ \hline
\rowcolor{gray!20} 
\cellcolor{gray!20}                      & Note-level Transformer  & 0.781                 & 0.777                 & 0.778                      & 0.631                 & 0.627                 & 0.628                  & 0.738                 & \textbf{0.734}        & \textbf{0.735}             & 0.476                        & 0.473                        & 0.474                           \\
\rowcolor{gray!20} 
\multirow{-2}{*}{\cellcolor{gray!20}HSD} & T3MS                    & \textbf{0.788}        & \textbf{0.778}        & \textbf{0.782}             & \textbf{0.682}        & \textbf{0.673}        & \textbf{0.677}         & \textbf{0.739}        & 0.731                 & 0.734                      & \textbf{0.517}               & \textbf{0.512}               & \textbf{0.514}                  \\ \hline
\end{tabular}
\end{table*}

Table \ref{table_note} presents the note-level transcription performance on ST500 and HSD, respectively.
T3MS exhibited superior performances in most metrics on the ST500 test set. Notably, even in the metrics where it did not outperform MusicYOLO-I, the performance difference between T3MS and MusicYOLO-I was no greater than 0.027. This maximum difference was observed in the experiments for the offset precision that focuses solely on the offset. Yet, MusicYOLO requires vocal separation, which means that it does not operate in a fully end-to-end manner like T3MS.

Interestingly, while MusicYOLO-I tended to have higher precision than recall, T3MS showed balanced precision and recall values. This suggests that MusicYOLO-I tends to miss many notes while rarely detecting notes that do not exist in the label, whereas T3MS demonstrates balanced presence between the missed and falsely detected notes. Consequently, T3MS achieved the highest F1 score among all the methods, thereby indicating its effectiveness in achieving the balance between the precision and recall.

T3MS showed better performance on the ST500 test set than the note-level Transformer. This improvement was achieved without the filtered DALI data, which improves the performance of the note-level Transformer by increasing the amount of training data. 
It suggests that a new approach of incorporating the recognition of the note values was effective for enhancing the performance of the note-level transcription that predicts the onset, offset, and pitch.

Additionally, we evaluated the note-level transcription performance on HSD that was solely used for evaluation. We compared the performances of T3MS and the note-level Transformer, as both models were utilized to evaluate the note value recognition performance. Overall, lower performance was observed on HSD compared to the ST500 test set, as indicated at the bottom of Table \ref{table_note} with a gray background. 
A notable decline was observed in the performances of the offset and note-level metrics compared to the other metrics.
The disparity can be attributed to the fact that the offset and note-level metrics consider the predicted offset time, suggesting that the performance drop originates from differences in the offset labeling methods between ST500 and HSD.

Results also demonstrated that T3MS generally outperformed the note-level Transformer across most metrics evaluated on HSD without the filtered DALI data. This suggests evidence for the effectiveness of the proposed approach. However, the note-level Transformer achieved higher recall and F1 scores when evaluating the onset time and pitches. Its proficiency in the pitch recognition demonstrates that it can handle various audio features more effectively, due to the additional use of the filtered DALI data during training.

\subsection{Comparison of Note Value Recognition Performances}

\begin{table*}
\centering
\caption{Comparison results for time-aligned note value recognition performances on the HSD test set}
\label{table_nl}
\begin{tabular}{c|ccc|ccccc|ccccc}
\hline
\multirow{2}{*}{model} & \multicolumn{3}{c|}{Onset + Pitch + Note Value}  & \multicolumn{5}{c|}{Onset + Offset + Note Value {(CN1)}}                             & \multicolumn{5}{c}{Onset + Offset + Pitch + Note Value {(CN2)}}                      \\
                       & P              & R              & F              & P              & R              & F              & acc            & MSE            & P              & R              & F              & acc            & MSE            \\ \hline
T3MS                   & \textbf{0.403} & \textbf{0.398} & \textbf{0.400} & \textbf{0.373} & \textbf{0.367} & \textbf{0.369} & \textbf{0.688} & \textbf{0.470} & \textbf{0.354} & \textbf{0.349} & \textbf{0.351} & \textbf{0.546} & \textbf{0.796} \\
Grid Quantization      & 0.382          & 0.379          & 0.380          & 0.331          & 0.328          & 0.329          & 0.668          & 0.493          & 0.318          & 0.315          & 0.316          & 0.517          & 0.856          \\
MuseScore              & 0.252          & 0.248          & 0.249          & 0.168          & 0.163          & 0.165          & 0.337          & 1.353          & 0.158          & 0.155          & 0.156          & 0.341          & 1.440          \\ \hline
\end{tabular}
\end{table*}

The note value recognition performance was evaluated by using 55 songs from the HSD test set. T3MS was compared with three methods: (1) the grid quantization using the transcription results from the note-level Transformer and madmom beat tracking results, (2) conversion of the transcription results of the note-level Transformer into a musical score using MuseScore\cite{musescore}, and (3) the transcription results of AnthemScore\cite{anthemscore}.
The first two methods enabled the evaluation of time-aligned note values by estimating the note values from the note-level transcription results. However, the last method using AnthemScore only extracted the pitches and note values, thereby limiting its evaluation to the symbolic domain metrics.
The results recognized by MuseScore and AnthemScore were exported in MusicXML\cite{musicxml} format. 
Anthemscore used vocal audio extracted by Spleeter\cite{spleeter} as an input since it transcribes all notes in audio, not just notes of a singing melody.

Table \ref{table_nl} shows the performances of recognizing the time-aligned note values. T3MS that directly extracts the note values from audio outperformed the other two methods of applying quantization after note-level transcription in all metrics. 
The precision, recall, and F1 scores of T3MS were improved by an average of 0.03 compared to the grid quantization method that achieved the second-best results across all metrics.
Since these metrics considered not only the note values but also the pitches and timing information, the accuracy and MSE calculated for correctly recognized notes should be considered.
Upon comparisons, T3MS demonstrated better performance with higher accuracy and lower MSE. Compared to the grid quantization, it demonstrated improvements of 0.029 in accuracy and 0.06 in MSE for CN2.

\begin{table}
\centering
\caption{Comparison results for error rates on the HSD test set}
\label{table_er}
\begin{tabular}{c|ccc|ccc}
\hline
\multirow{2}{*}{model} & \multicolumn{3}{c|}{note only}                & \multicolumn{3}{c}{note and rest}             \\
                  & NER           & PER           & VER           & NER           & PER           & VER           \\ \hline
T3MS              & \textbf{64.4} & 11.5          & \textbf{46.4} & \textbf{78.6} & \textbf{28.8} & \textbf{60.0} \\
Grid Quantization & 65.3          & \textbf{10.3} & 48.2          & 88.0          & 36.3          & 69.6          \\
MuseScore         & 94.7          & 10.9          & 75.9          & 114.8         & 36.5          & 94.1          \\
AnthemScore       & 129.4         & 55.4          & 65.2          & 149.1         & 81.5          & 85.8          \\ \hline
\end{tabular}
\end{table}

The note value recognition performance can also be evaluated in the symbolic domain using the error rate metrics that do not consider  the onset and offset times. In this evaluation, we can compare the recognition results of AnthemScore that extracts the pitches and note values from vocal audio with the other models. To calculate the error rate, we converted each recognition result into a word sequence in which each word represents the pitch or note value. The error rates were evaluated in two cases: one that considers only notes, while the other that considers both notes and rests. 
T3MS achieved the lowest error rates in all three metrics when considering both notes and rests, and the lowest NER and VER values when considering notes only, as shown in Table \ref{table_er}. However, the grid quantization and MuseScore yielded lower PER values than T3MS when considering notes only. This can be attributed to the strength of the note-level Transformer in the pitch recognition. AnthemScore exhibited high error rates due to the inaccuracies in vocal separation such as detection of instrument sound as vocal or failure to recognize a singing melody.

\subsection{Qualitative Evaluation}
\label{section_qual_eval}
\begin{figure}[!t]
    \centering
    \includegraphics[width=\linewidth]{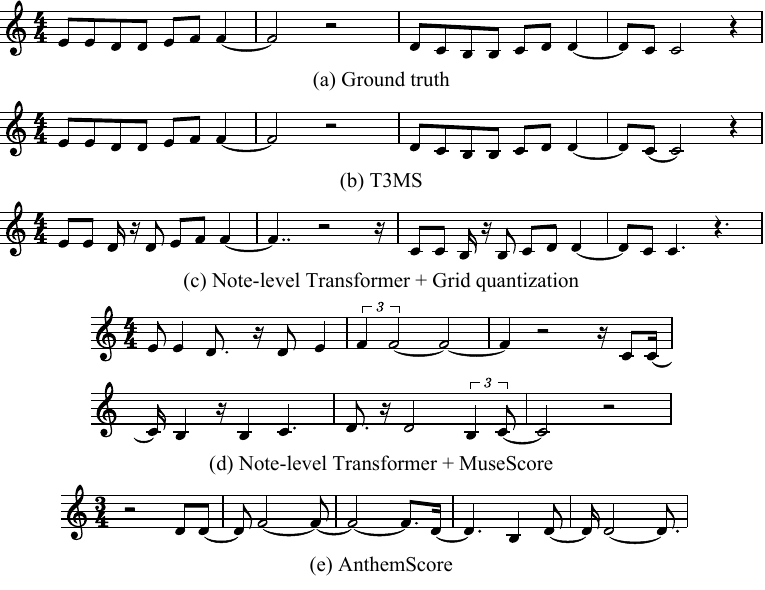}
    \caption{Visualization of the transcription results as a musical score. The first figure (a) represents the ground-truth label. The following figures show the results of (b) T3MS, (c) grid quantization with the note-level Transformer, (d) MuseScore with the note-level Transformer, and (e) AnthemScore from top to bottom.}
    \label{fig_vis}
\end{figure}
To provide a deeper understanding of the transcription performance, the results were visualized in the form of lead sheets and analyzed qualitatively.
Fig. \ref{fig_vis} shows lead sheets for the ground-truth label and the recognition results of the four approaches.
The recognition result of T3MS closely matched the ground-truth, except for the note tie.
The visualization result of the grid quantization method was comparable to the ground-truth but included incorrect note values, such as the notes with double dots.
MuseScore produced a lead sheet more complex than the ground-truth, characterized by triplet notes. In contrast, AnthemScore extracted a simpler lead sheet with fewer notes and longer note values for each note. Notably, AnthemScore was able to recognize a time signature other than 4/4, which was not possible for the other methods.

\begin{figure}[!t]
    \centering
    \includegraphics[width=\linewidth]{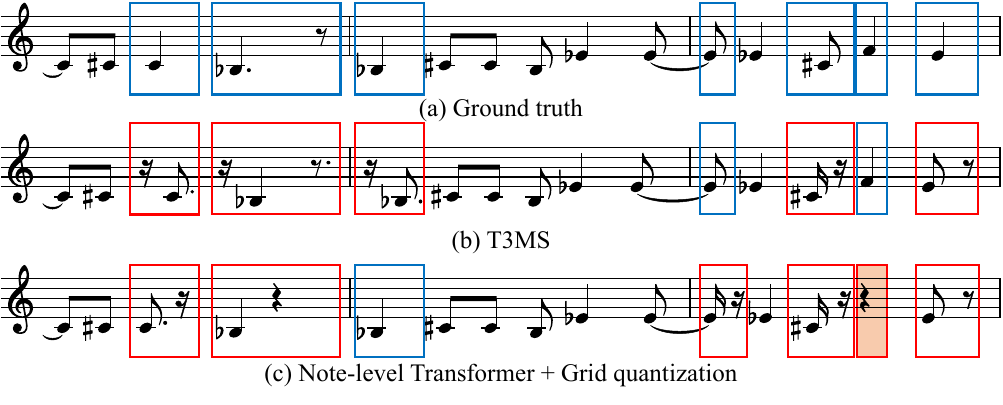}
    \caption{Comparisons of the transcription results on HSD. Red boxes represent parts of the score with errors, while blue boxes indicate the corresponding ground-truth or correct prediction results. A red box with a background indicates the missed note, which is not related to the offset prediction.}
    \label{fig_hsd}
\end{figure}

In Section \ref{section_hsd}, we analyzed the difference between ST500 for training and HSD for evaluation.
The offset time predicted by the model trained on ST500 was positioned ahead of the offset label of HSD, resulting in more frequent rests and shorter note values.
Fig. \ref{fig_hsd} presents the transcription results on HSD to analyze the difference of the offset time. We compared the ground-truth with the results from T3MS and the grid quantization method since these two methods outperformed the other models, as shown in Fig. \ref{fig_vis}.
The recognition results of these two models showed more frequent rests or rests with longer note values compared to the ground-truth. The first and third boxes of each result in Fig. \ref{fig_hsd} illustrate these examples, respectively. Additionally, the grid quantization method failed to detect the note, which is not due to the different offset labeling method of HSD, as depicted as the red box with a background in Fig. \ref{fig_hsd}c.

\begin{figure}[!t]
    \centering
    \includegraphics[width=\linewidth]{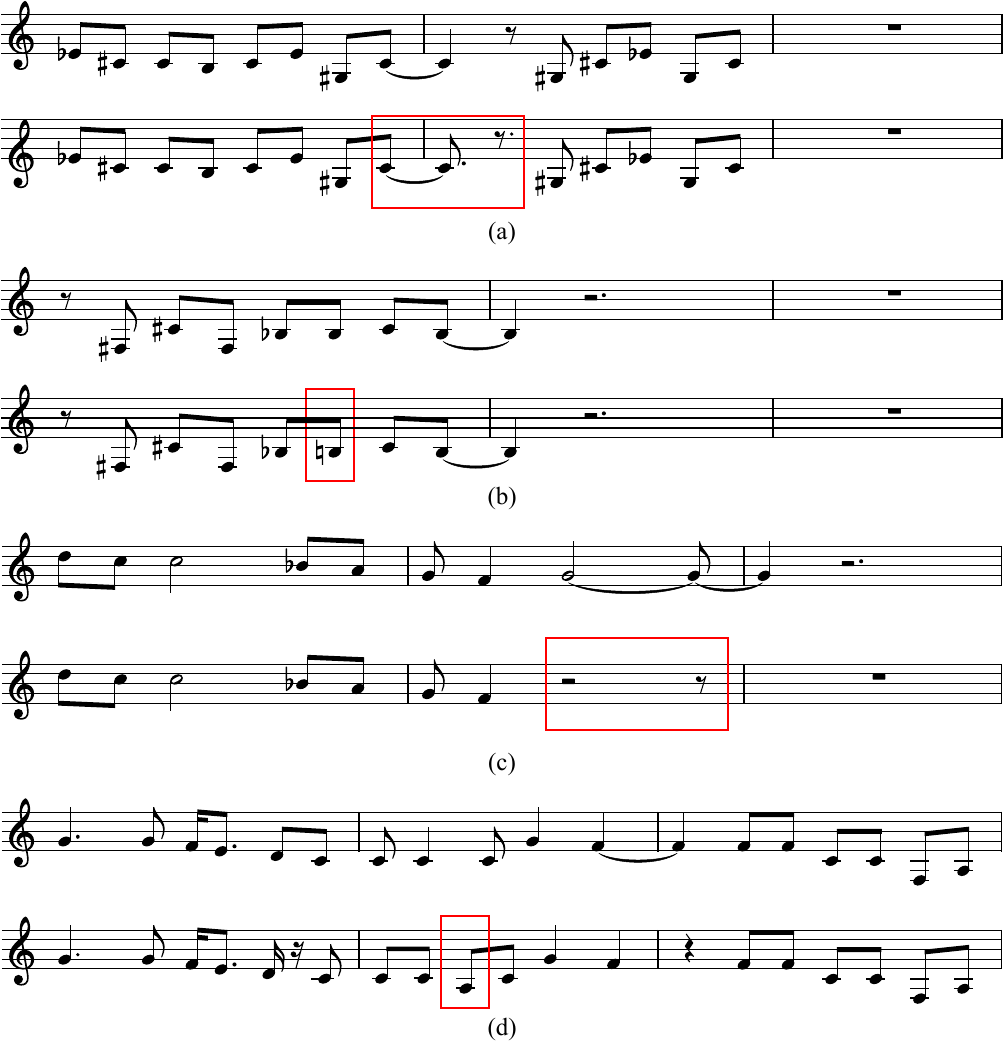}
    \caption{Error cases of T3MS: (a) incorrect prediction of the note value, (b) incorrect prediction of the pitch, (c) failure to detect the note, and (d) recognition of the extra notes. Each subfigure illustrates the ground-truth at the top and the recognition results by T3MS at the bottom. Prediction results that differ from the label are highlighted in red boxes.}
    \label{fig_error}
\end{figure}

Fig. \ref{fig_error} illustrates the examples of errors that may occur in the recognition results of T3MS. The error cases depicted in Fig. \ref{fig_error}b, Fig. \ref{fig_error}c, and Fig. \ref{fig_error}d can be observed in the results of the previous note-level transcription models. 
The errors associated with the note value recognition in Fig. \ref{fig_error}a can be divided into cases where only the note value is incorrectly recognized and cases where the onset or offset times are also incorrectly recognized. 

As the note value is closely related to the note onset and offset times, errors in the note-level transcription may also affect the note value recognition. Such errors are often observed when vocalist sings in a whisper-like manner or when other instruments or voices interfere with the main vocal. 
These factors can also affect the note-level singing melody transcription performance, highlighting the need for robust models capable of effectively handling such confusing inputs.

\section{Conclusion}
This paper proposed an end-to-end framework for a new task, named the extended note-level singing melody transcription, to generate time-aligned musical scores directly from audio. We introduced a pseudo-labeling method that generates pseudo-labels from the datasets for the note-level transcription to overcome the lack of the note value labels. The proposed model outperformed the state-of-the-art note-level singing melody transcription models and achieved improved note value recognition performances compared to the other existing models. Qualitative evaluation further confirmed the superior performance of the proposed model in predicting the time-aligned note values.

Although the proposed model demonstrated the capability to generate more accurate time-aligned lead sheets in a single training process, it still has some limitations, such as the inability to recognize triplet notes or time signatures other than 4/4. Recognizing triplet notes necessitates reducing the note value unit by a factor of three, thereby requiring an adjustment of the pseudo-labeling process and the model retraining within the proposed framework. To accommodate various time signatures, it is necessary to estimate rhythmic structure additionally. Furthermore, more robust model is necessary to handle various musical audio types, including loud accompaniments, multi-vocal, and whisper-like singing styles. Sufficient note value labels with the onset, offset, and pitch information are also required for supervised learning.

Future work will concentrate on accommodation of various time signatures and note values, in addition to collection of large datasets with precise note value labels. This will facilitate the generation of more accurate time-aligned scores, even in cases with diverse musical styles and accompaniments. The time-aligned scores produced by the extended note-level transcription in such cases will enable various potential applications in music education, automatic music transcription, and musicological analysis.

\section*{Acknowledgments}
This work was supported by the National Research Foundation of Korea (NRF) grant funded by the Korea government (MSIT) (No. NRF-2019R1F1A1053366).

\bibliographystyle{IEEEtran}
\bibliography{IEEEabrv, ref}
 
\vspace{11pt}

\vspace{-33pt}
\begin{IEEEbiography}[{\includegraphics[width=1in,height=1.25in,clip,keepaspectratio]{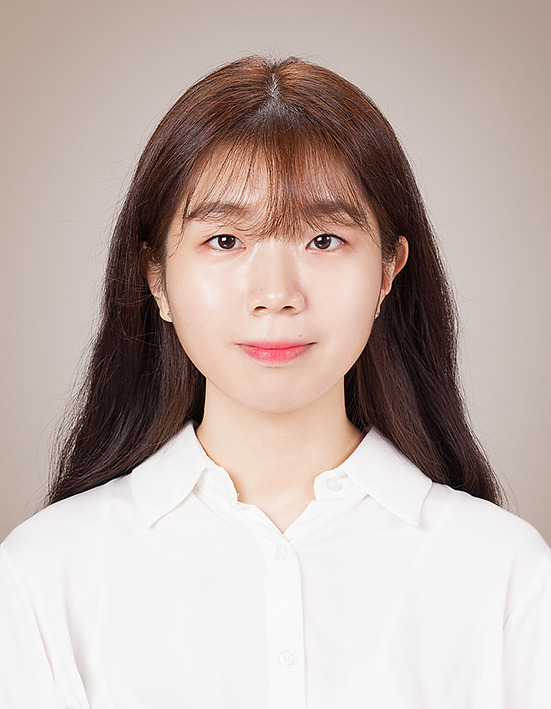}}]{Leekyung Kim}
received the B.S. degree in industrial engineering from Seoul National University (SNU), Korea, in 2021. She is currently pursuing the Ph.D. degree with the Information Management Laboratory, Department of Industrial Engineering from Seoul National University (SNU). 
Her current research interests include automatic music transcription, audio signal processing, and deep learning applications.
\end{IEEEbiography}
\begin{IEEEbiography}[{\includegraphics[width=1in,height=1.25in,clip,keepaspectratio]{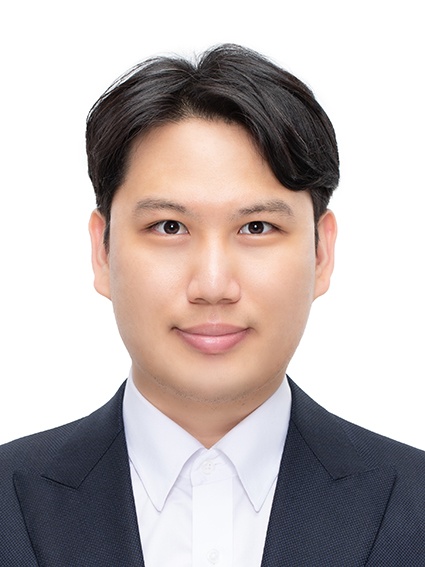}}]{Sungwook Jeon}
received the B.S. degree in industrial engineering from Seoul National University (SNU), Korea, in 2012. 
He is currently pursuing the Ph.D. degree with the Information Management Laboratory, Department of Industrial Engineering from Seoul National University (SNU). 
His current research interests include generative model for music and ai chatbot.
\end{IEEEbiography}
\begin{IEEEbiography}[{\includegraphics[width=1in,height=1.25in,clip,keepaspectratio]{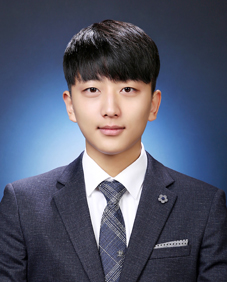}}]{Wan Heo}
received the B.S. degree in industrial engineering from Seoul National University (SNU), Korea, in 2017. 
He is currently pursuing the Ph.D. degree with the Information Management Laboratory, Department of Industrial Engineering from Seoul National University (SNU). 
His current research interests include symbolic music generation and editing, automatic music transcription, and deep learning applications using sequential and tabular data.
\end{IEEEbiography}
\begin{IEEEbiography}[{\includegraphics[width=1in,height=1.25in,clip,keepaspectratio]{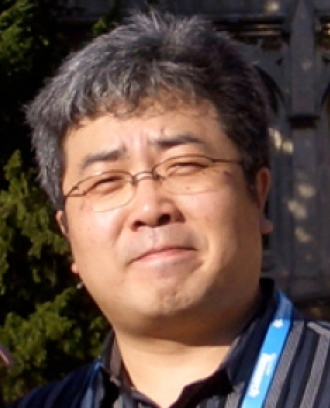}}]{Jonghun Park}
received the Ph.D degree in industrial and systems engineering with a minor in computer science from the Georgia Institute of Technology, Atlanta, in 2000. He is currently a professor in the Department of Industrial Engineering, Seoul National University (SNU), Korea. 
His research interests include generative artificial intelligence and deep learning applications.
\end{IEEEbiography}

\vspace{11pt}

\vfill

\end{document}